# Tomographic Image Reconstruction Using an Advanced Score Function (ADSF)

Wenxiang Cong, Wenjun Xia, Ge Wang

Biomedical Imaging Center, Center for Computational Innovations, Center for Biotechnology and Interdisciplinary Studies

Department of Biomedical Engineering,

Rensselaer Polytechnic Institute, Troy, NY 12180

**Abstract:** Computed tomography (CT) reconstructs volumetric images using X-ray projection data acquired from multiple angles around an object. For low-dose or sparse-view CT scans, the classic image reconstruction algorithms often produce severe noise and artifacts. To address this issue, we develop a novel iterative image reconstruction method based on maximum a posteriori (MAP) estimation. In the MAP framework, the score function, i.e., the gradient of the logarithmic probability density distribution, plays a crucial role as an image prior in the iterative image reconstruction process. By leveraging the Gaussian mixture model, we derive a novel score matching formula to establish an advanced score function (ADSF) through deep learning. Integrating the new ADSF into the image reconstruction process, a new ADSF iterative reconstruction method is developed to improve image reconstruction quality. The convergence of the ADSF iterative reconstruction algorithm is proven through mathematical analysis. The performance of the ADSF reconstruction method is also evaluated on both public medical image datasets and clinical raw CT datasets. Our results show that the ADSF reconstruction method can achieve better denoising and deblurring effects than the state-of-the-art reconstruction methods, showing excellent generalizability and stability.

X-ray computed tomography (CT) is the most popular imaging modality used in various fields, including medical imaging, homeland security, and industrial applications [1]. Filtered backprojection (FBP) is an analytical method for tomographic image reconstruction [2], and often produces severe noise and artifacts in low-dose or sparse-view settings [3]. To address this challenge, iterative reconstruction methods were developed to incorporate the photons statistics and desirable image prior models especially sparsity [3-5]. The key issue in this iterative reconstruction method is the image prior modeling. A representative method based on compressed sensing (CS) converts images into sparse data and then performs image reconstruction through sparse regularization such as total variation (TV) minimization [4]. However, CS-based image reconstruction tends to over smoothen textures and eliminate subtle details in a reconstructed image. To overcome this shortcoming, low dimensional manifold model (LDMM) was developed for image reconstruction utilizing the low dimensional characteristics of the image patch manifold [6, 7]. However, these presumptions did not accurately and comprehensively reflect actual image structures, especially for sophisticated biomedical images.

Deep learning is a powerful approach to perform various types of uncertainty estimation and data modeling [8]. Supervised learning was applied to perform image denoising and deburring by establishing a convolution neural network-based image processing model [9], and implement image reconstruction by replacing regularization terms with a trained convolutional neural network (CNN) within the framework of an unrolling iterative optimization scheme [10, 11]. The supervised learning requires a labeled dataset pairing each input sample with the corresponding target, which is usually unavailable in practice. Recently, diffusion models or score matching models have attracted a major attention for image generation [12] and image reconstruction [13, 14]. In an unsupervised fashion, the score matching methods learn a score function, i.e., the gradient of the logarithmic probability density function, from a training data set without any labels [13]. In contrast to conventional regularized reconstruction methods that aim to constrain images to become sparse, low-rank, and low-dimensional, the scoring function can generate an optimal image prior using the data-driven approach, allowing an effective representation of the image distribution.

In this paper, we model the number of X-ray photon reaching each pixel on the detector as a Poisson distribution, and perform CT image reconstruction using maximum a posteriori (MAP) estimation. In the MAP framework, the score function plays a crucial role as an image prior in the iterative image reconstruction process. By leveraging Gaussian mixture model to characterize the discrepancy between a reconstructed image and the target image, we derive a novel score matching formula to generate an advanced score function (ADSF) through deep learning. By integrating a new ADSF into the image reconstruction process, we develop a new iterative reconstruction method for image reconstruction. The convergence of the iterative reconstruction algorithm is proven through mathematical analysis. The performance of proposed ADSF reconstruction method is also evaluated on both public medical image datasets and clinical raw datasets, demonstrating the superiority of our approach over state-of-the-art image reconstruction methods in terms of accuracy and generalizability.

## Results

Based on the benchmark dataset used in the NIH-AAPM-Mayo Clinic low-dose CT Challenge, the advanced score function (ADSF) was established through deep learning according to the new score matching formula Eq. (16) in the Methods section. Using the obtained ADSF, we conducted extensive experiments on public medical image datasets and clinical CT raw datasets to evaluate the performance of the ADSF reconstruction method in comparison with the representative image reconstruction methods, including the filtered backprojection (FBP) [1], model-based iterative reconstruction (MBIR) with total variation minimization (MBIR-TV) [4, 15], and state-of-the-art score function based image reconstruction (SFrecon) [13, 14].

**Phantoms**

CT images from the NIH-AAPM-Mayo Clinic CT Grand Challenge were selected as realistic digital phantoms to evaluate the performance of the ADSF reconstruction method. Specifically, the dataset contains 2,378 images from 10 patients. Image phantoms were randomly selected from 2,378 images. In the X-ray imaging simulation of the image phantoms, the scan trajectory radius was 595mm. The distance from the source to the detector was 1085.6mm. There were 736 detector elements with 1.2858mm pitch, equiangularly distributed along a curvilinear array. Fan-beam projections of X-ray imaging were generated using an industrial CT simulator called CatSim at 120kV and 100mAs. A total of 360 projection views were uniformly acquired over a 360 degree range. We performed fan-beam image reconstruction for 100 digital phantoms using FBP, MBIR-TV, SFrecon, and our ADSF reconstruction methods, respectively. The reconstructed images were quantitatively evaluated using peak signal-to-noise ratio (PSNR) and structural similarity index (SSIM) metrics. **Table 1** presents the quantitative results of the reconstructed images with respect to PSNR and SSIM, showing that the ADSF reconstruction method achieved higher average PSNR and SSIM values than other competing reconstruction methods. **Figure 1** displays the representative images and zoom-in patches for visualization. By comparing reconstructed images, it can be observed that the image reconstructed using the FBP is noisier than the image reconstructed by the other reconstruction methods in the low-dose and few-view scenarios, while the ADSF reconstruction method achieves excellent denoising and deblurring effects and well preserves structural information. The qualitative and quantitative results show that the proposed ADSF reconstruction method outperforms the other reconstruction methods.

*Table 1. Quantitative comparison of reconstructed image quality*

| Metric | FBP | MBIR-TV | SFrecon | ADSF |
|---|---|---|---|---|
| SSIM | 0.9433±0.0214 | 0.8958±0.028 | 0.8608±0.0238 | 0.9602±0.0195 |
| PSNR | 40.0317±1.956 | 38.5586±1.597 | 37.0325±1.743 | 41.7077±1.348 |

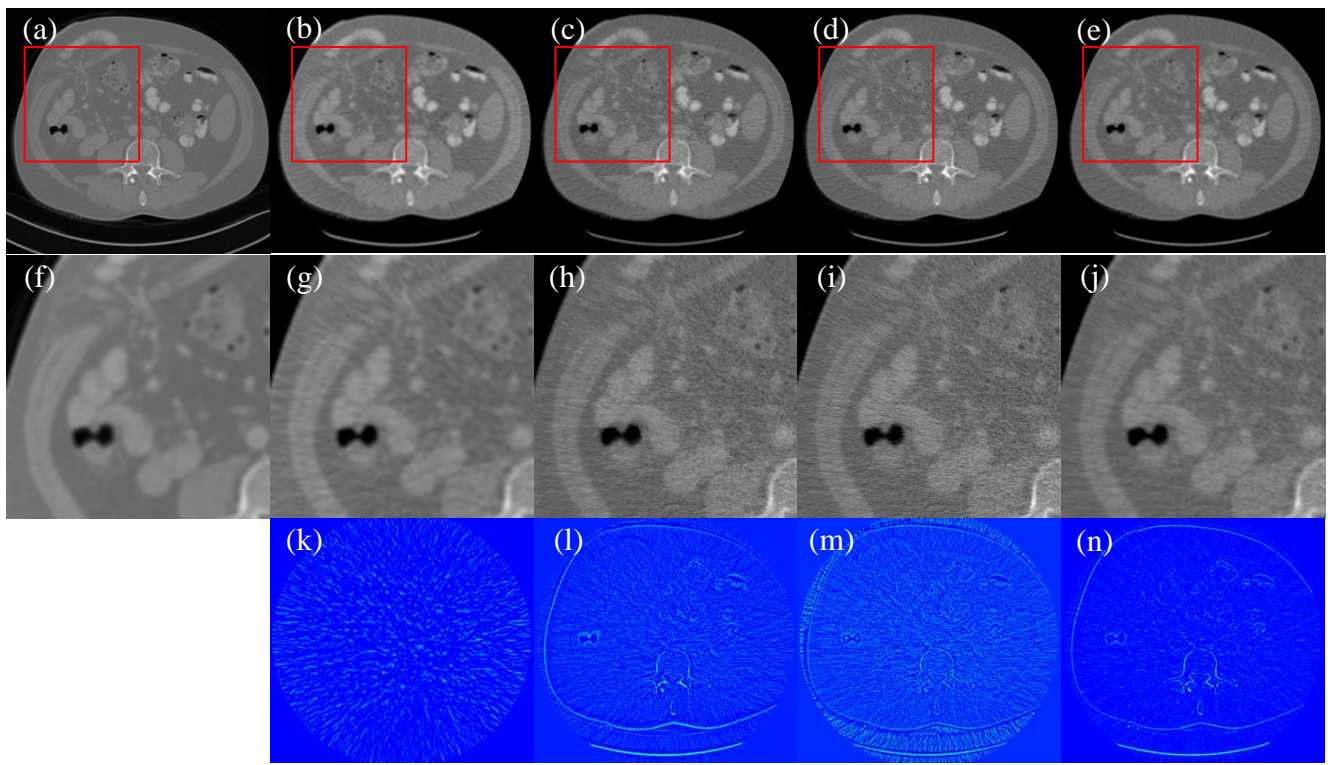

*Figure 1. Image reconstructions using competing algorithms. (a) The ground truth image, (b) the image reconstructed using FBP, (c) the image reconstructed using MBIR-TV, (d) the image reconstructed using SFrecon, (e) the image reconstructed by ADSF reconstruction method, and (f)-(j) the enlarged ROIs defined by the red box in images (a-e), respectively. (k)-(n) The error images of (b)-(e) relative to the ground truth image (a). The display window is [-800, 800] HU.*

**Clinical CT Raw Datasets**

**Clinical raw data from the Siemens scanner:** NIH-AAPM-Mayo Clinic Low-dose CT Grand Challenge contained 10 anonymized patient normal dose abdominal CT images. The dataset was acquired on the Siemens SOMATOM Definition Flash CT system with voltage 120 kV and 200 mAs. The scanner had a scanning radius of 595mm. The distance from the source to the detector was 1085.6mm. There were 736 detector elements with 1.2858mm pitch, equiangularly distributed along a curvilinear array. Over a 360 degree range, 768 projection views were uniformly acquired for image reconstruction. Here, the helical raw data were converted to flat detector fan-beam projections. Based on the fan-beam projection dataset, we performed 100 image reconstructions from one-third projection views using FBP, MBIR-TV, SFrecon, and ADSF reconstruction methods, respectively. **Table 2** shows the quantitative results of the reconstructed images relative to the reference image reconstructed from the 2304 full-view projections using FBP. **Figure 2** displays the representative images, showing better performance in terms of noise removal as well as preservation of weak edges and exhibit structural information.

*Table 2. Quantitative indexes of reconstructed image quality on Siemens raw dataset*

| Metric | FBP | MBIR-TV | SFrecon | ADSF |
|---|---|---|---|---|
| SSIM | 0.9035±0.0203 | 0.9182±0.024 | 0.7874±0.0208 | 0.9480±0.0198 |
| PSNR | 40.6499±1.896 | 39.1437±1.696 | 32.5529±1.773 | 43.7712±1.548 |

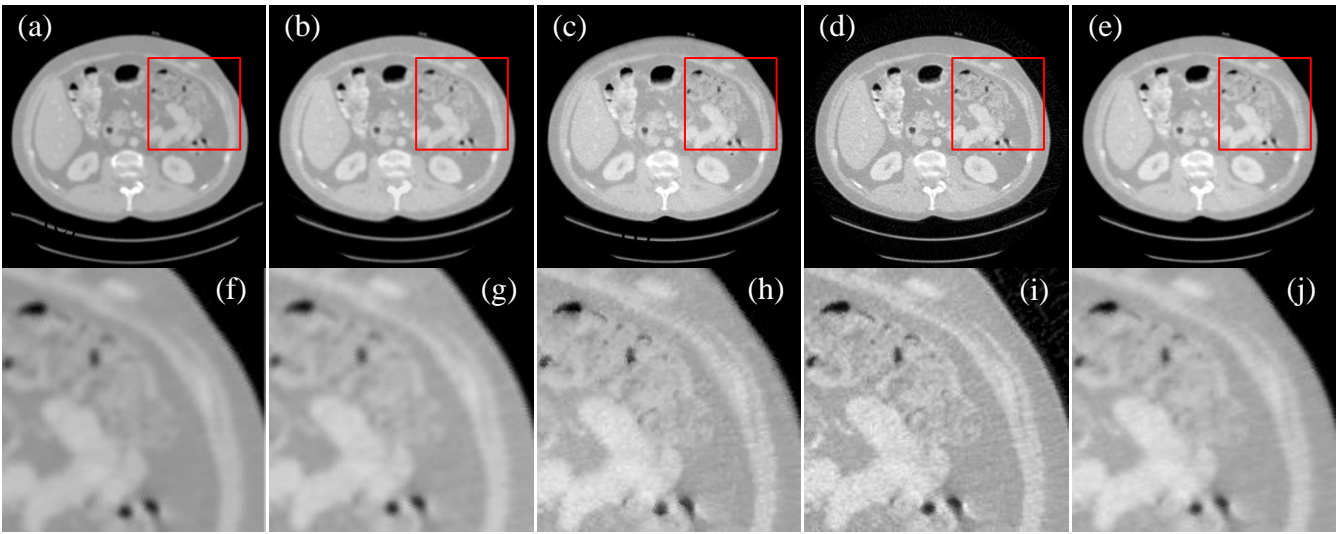

*Figure 2.* CT image reconstructions from raw CT data generated on the Siemens scanner. (a) The image reconstructed from 2304 projections using FBP, (b) the image reconstructed from 768 projections of using FBP, (c) the image reconstructed by MBIR-TV, (d) the image reconstructed by SFrecon, (e) the image reconstructed by ADSF reconstruction method, and (f)-(j) the enlarged ROIs defined by the red box in (a)-(e), respectively. The display window is [-800, 800] HU.

**Clinical raw data from the GE scanner:** Furthermore, a clinical raw dataset were also acquired from GE CT scanner. The scanner had a scanning radius of 541mm. The distance from source to detector was 949mm. There were 888 detector elements with 1.024mm pitch, equiangularly distributed along a curvilinear array. 984 projection views were uniformly acquired over a 360 degree range. Here, we uniformly selected 328 projection views for few-view image reconstruction. **Table 3** gives a quantitative comparison of the reconstructed images relative to the reference image reconstructed from 984 full-view projections using FBP. From image reconstruction results shown **in Figure 3**, it is clear that FBP, MBIR-TV, and SFrecon methods still exhibit blurry boundaries and unremoved noise, while our ADSF reconstruction method is quite robust in denoising and deblurring.

*Table 3. Quantitative indexes of reconstructed image quality on GE raw dataset*

| *Metric* | **FBP** | **MBIR-TV** | **SFrecon** | **ADSF** |
|---|---|---|---|---|
| *SSIM* | *0.8201* | *0.9160* | *0.7395* | *0.9610* |
| *PSNR* | *34.1065* | *40.4881* | *33.4690* | *46.0023* |

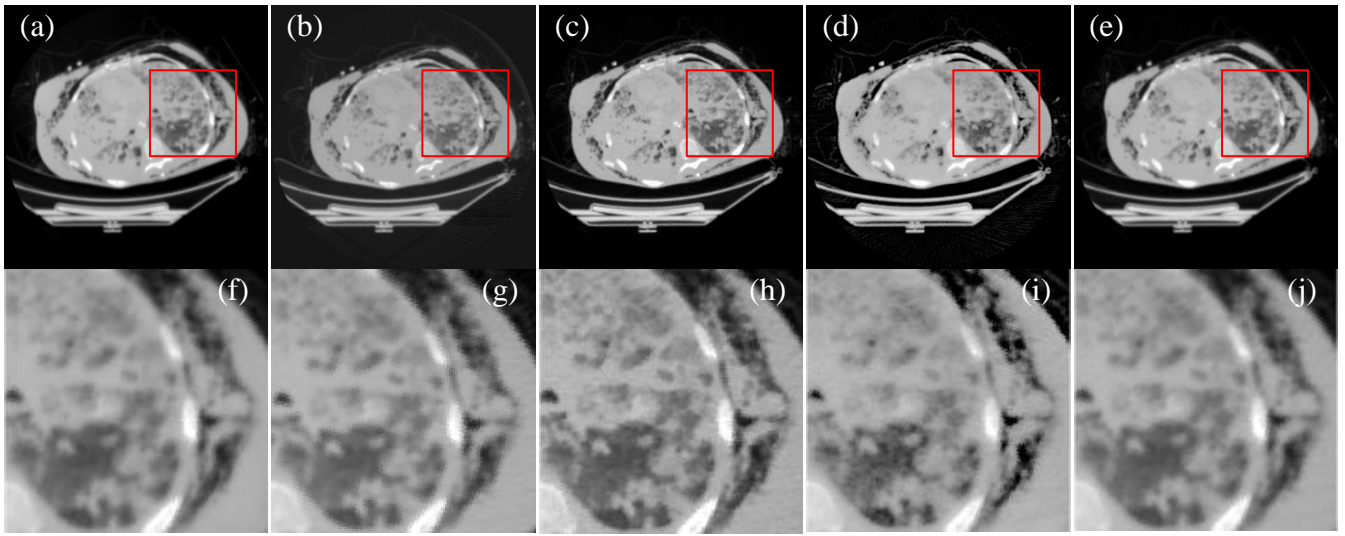

*Figure 3. CT image reconstructions from raw data generated by GE scanner. (a) The image reconstructed from 984 projections using FBP, (b) the image reconstructed from 328 projections using FBP, (c) the image reconstructed by MBIR-TV, (d) the image reconstructed using SFrecon, (e) the image reconstructed by ADSF reconstruction method, and (f)-(j) the enlarged ROIs in the red box as shown in (a)-(e). The display window is [-900, 400] HU.*

## Discussion

Based on X-ray imaging physics, the number of photons reaching individual pixels on the detector is model by a Poisson distribution, and CT image reconstruction is performed via maximum a posteriori (MAP) estimation. Under the MAP estimation framework, the CT images can be reconstructed by maximizing likelihood of the Poisson statistics and an image prior probability density. The logarithmic probability density distribution as image prior plays a crucial role in the MAP estimation. Hence, image prior modeling is an important task in iterative reconstruction methods. Generally, there is no explicit formula for an image prior model. Conventional iterative reconstruction methods usually utilize regularization techniques such as compressive sensing to represent the image prior model, which assumes that images can be converted into sparse, low-rank, or low-dimensional counterparts using transformation methods such as total variation, wavelet transform, or Fourier transform. However, these presumptions do not always accurately reflect structure and characteristics of actual images, especially for sophisticated biomedical images.

Nevertheless, the emerging machine learning provides a powerful technique for the data modeling. Based on the score matching model [13, 16], the score function, i.e., the gradient of the logarithmic probability density distribution, can be established as the image prior through deep learning. The score function serves as the key element in this iterative process, contributing to image refinement in the each iteration. By leveraging the Gaussian mixture model to characterize noise distributions between a reconstructed image and the target image, we have developed a novel score matching formula to learn an advanced score function (ADSF) from a CT image dataset. Because the image reconstruction has a complex relationship between the reconstructed images and the true images, the Gaussian mixture model can describe a practical image noise distribution much better than the popular single Gaussian distribution. The new score function learns directly from the data distribution, and provides an advanced mechanism for effective and efficient extraction of data-driven prior information and a superior representation of underlying images for image reconstruction.

In summary, we have proposed a novel score-matching formula to generate an advanced score function (ADSF) through deep learning from an image dataset. By incorporating the ADSF into the image reconstruction process, a new ADSF iterative reconstruction method has been developed in the MAP estimation framework to improve image reconstruction quality. The convergence of the ADSF iterative reconstruction algorithm has been proved through mathematical analysis. The performance of ADSF image reconstruction method has been evaluated on both public medical image datasets and clinical raw datasets. By comparing with the competing image reconstruction techniques such as the filtered backprojection (FBP), model-based iterative reconstruction (MBIR) with total variation minimization (MBIR-TV), and the state-of-the-art score function-based reconstruction (SFrecon) method, the quantitative results has shown that our proposed ADSF reconstruction method can achieve higher quality images in terms of PSNR and SSIM metric on diverse dataset. For Siemens and GE clinical CT raw datasets, the ADSF image reconstruction approach has achieved better denoising and deblurring effects than the competing methods, showing excellent generalizability and stability. The proposed new score matching formula can also be applied to image denoising, image deblurring, and image generation. Further algorithmic optimization and systematic evaluation are in progress to translate this new approach into clinical applications.

## Methods

### Image reconstruction approach

In X-ray imaging, the number of X-ray photons recorded by a detector element is a random variable $\xi$, which can be modeled as the Poisson distribution [2]:

$$p(\xi = y_i) = \frac{(\bar{y}_i)^{y_i}}{y_i!} \exp(-\bar{y}_i), \tag{1}$$

where $\bar{y}_i$ is the expectation value of recorded X-ray photons along a path $l$ from the X-ray source to the $i^{th}$ detector element, and obeys the Beer-Lambert law:

$$\bar{y}_i = n_i \exp\left(-\int_l \mu(\vec{r}) dl\right), \tag{2}$$

where $n_i$ is the number of X-ray photons recorded by the $i^{th}$ detector element in the blank scan, without any object in the beam path, and $\mu(r)$ is the linear attenuation coefficient distribution within an object to be reconstructed. For the numerical implementation, Eq. (2) is discretized as,

$$\bar{y}_i = n_i \exp(-A_i \mu), \tag{3}$$

where $\mu$ is a vector of pixel values in the linear attenuation coefficient image, and $A_i$ is weighting coefficients of the pixel values along the $i^{th}$ beam path. Assuming that measurements are independent, the likelihood function of the X-ray projection data can be obtained by

$$p(Y|\mu) = \prod_{i=1}^{m} \frac{(\bar{y}_i)^{y_i}}{y_i!} \exp(-\bar{y}_i), \tag{4}$$

where $Y = (y_1, y_2, \cdots, y_m)^T$ is the number of photons measured by detectors, and $m$ is the total number of X-ray detectors. Based on the Bayesian theorem: $p(\mu|Y)p(Y) = p(Y|\mu)p(\mu)$, the image reconstruction can be performed using the maximum a posteriori (MAP) estimation, which is equivalent to the following minimization problem [17]:

$$\mu_{min} = \arg\min_{\mu} \left( \sum_{i=1}^{m} [\bar{y}_i - y_i \log(\bar{y}_i)] - \log p(\mu) \right), \tag{5}$$

where $\log p(\mu)$ is the logarithmic probability density of an attenuation image, which expresses the prior knowledge about the underlying images. Combining Eqs. (3)-(5), we have

$$\mu_{min} = \arg\min_{\mu} \left( \sum_{i=1}^{m} [n_i \exp(-A_i \mu) + y_i A_i \mu] - \log p(\mu) \right). \tag{6}$$

Applying the second-order Taylor approximation, Eq. (6) can be simplified to the following quadratic optimization problem [4]:

$$\mu_{min} = \arg\min_{\mu} \left[ \frac{1}{2}(A\mu - b)^T D(A\mu - b) - \log p(\mu) \right] \tag{7}$$

where $A$ is the $m \times n$ system matrix composed of the row vectors $A_1, A_2, \cdots, A_m$, and $D$ is the diagonal matrix in the form $diag(y_1, y_2, \cdots, y_m)$. The optimization problem defined by Eq. (7) can be solved using the gradient-based method iteratively:

$$\mu^{k+1} = \mu^k - \left(\omega A^T D(A\mu^k - b) - \sigma \nabla \log p(\mu^k)\right), \quad k = 1, 2, \cdots \tag{8}$$

where $\nabla \log p(\mu)$ is the score function, defined as the gradient of the logarithmic probability density distribution with respect to the current image, and $\omega$ and $\sigma$ are parameters for trade-offs between data fidelity and image prior information.

## Convergence of score-function-based image reconstruction

For the convergent analysis of the iteration scheme Eq. (8), we assume that $A$ is a $m \times n$ system matrix of rank $n$. The probability density function $p(\mu)$ is assumed to be sufficiently smooth, and the Hessian matrix $\nabla^2 \log p(\mu)$ of the logarithmic probability density function has bounded eigenvalues, denoted by $|h_i| < C, \ i = 1, 2, \cdots, n$ [18, 19]. Based on these assumptions, we have the following Lemma for the convergence of the iteration scheme Eq. (8).

**Lemma 1:** The iteration scheme Eq. (8) is convergent.

**Proof**: Based on the iteration procedure Eq. (8), we obtain

$$\mu^{k+1} - \mu^k = (I - \omega A^T DA)(\mu^k - \mu^{k-1}) + \sigma \left[ \nabla \log p(\mu^k) - \nabla \log p(\mu^{k-1}) \right], \tag{9}$$

Based on the mean value theorem of a multivariate function, there exists a vector $\xi$ such that,

$$\log p(\mu^k) - \log p(\mu^{k-1}) = \nabla \log p(\xi)(\mu^k - \mu^{k-1}). \tag{10}$$

From Eqs. (9-10), we obtain

$$\mu^{k+1} - \mu^k = \left(I - \omega A^T DA + \sigma \nabla^2 \log p(\xi)\right)(\mu^k - \mu^{k-1}), \tag{11}$$

where $\nabla^2 \log p(\xi)$ is the Hessian matrix of the logarithmic probability density function and is symmetric. Since $(Ax)^T DAx = \sum_{i=1}^{m} y_i (A_i \cdot x)^2 > 0$ for any nonzero vector $x \in R^n$, the matrix $A^T DA$ is positive definite, denoting its smallest and largest eigenvalues as $\lambda_{\min}$ and $\lambda_{\max}$ respectively. From Eq. (11), it is easy to find that by choosing the parameters $\omega$ and $\sigma$ to satisfy that $0 < \sigma < \omega r_{\min}/C$ and $\sigma C/r_{\min} < \omega < (2 - \sigma C)/r_{\max}$, there exist a positive constant $0 < q < 1$ such that $\|\mu^{k+1} - \mu^k\| \leq q \|\mu^k - \mu^{k-1}\|$, and $\|\mu^{k+1} - \mu^k\| \leq q^k \|\mu^1 - \mu^0\|$. Hence, according to the Cauchy convergence criterion, the image sequence $\{\mu^k \mid k = 0, 1, \cdots\}$ must converge.

## Score function estimation

The score function is the gradient of the logarithmic probability density with respect to a current image, i.e., $\nabla \log p_{data}(x)$. The unknown score function in Eq. (8) must be estimated in advance to perform the image reconstruction. If a known dataset sampled from a data distribution $p_{data}(x)$ is available, the score function $\nabla \log p_{data}(x)$ can be estimated using the score matching method, which is introduced by Hyvärinen (2005) [20]. The score matching method is to train a neural network model $s_\theta(x)$ by optimizing model parameters $\theta$ to best match the score function $\nabla \log p_{data}(x)$. The task can be performed by minimizing the following objective function:

$$E_{p_{data}}\left[\left\|s_\theta(x) - \nabla \log p_{data}(x)\right\|_2^2\right]. \tag{12}$$

It has been demonstrated that the above objective function is equivalent to the following expectation [16]:

$$E_{p(\tilde{x}|x)p_{data}(x)}\left[\left\|s_\theta(\tilde{x}) - \nabla_{\tilde{x}} \log p(\tilde{x}|x)\right\|_2^2\right]. \tag{13}$$

The data $x$ is perturbed with a noise distribution $p(\tilde{x}|x)$, which can be accurately modeled by the Gaussian mixture,

$$p(\tilde{x}|x) = \int p(\sigma) G(\tilde{x}; x, \sigma^2 I) d\sigma, \tag{14}$$

where $G(\tilde{x}; x, \sigma^2 I)$ is the Gaussian distribution with the variance $\sigma^2$. The Gaussian mixture model represents complex noise effectively in real scenarios [21]. From Eq. (14), the gradient of the logarithmic probability density distribution can be calculated,

$$\begin{cases} \nabla_{\tilde{x}} \log p(\tilde{x}|x) = \int \lambda(\sigma, \tilde{x} - x) \frac{x - \tilde{x}}{\sigma^2} d\sigma \\ \lambda(\sigma, \tilde{x} - x) = \dfrac{p(\sigma) G(\tilde{x}; x, \sigma^2 I)}{\int p(\sigma) G(\tilde{x}; x, \sigma^2 I) d\sigma} \end{cases} \tag{15}$$

From Eqs. (13) and (15), the score matching objective function in Eq. (13) is simplified to the following objective function:

$$E_{p(\sigma)} E_{p(\tilde{x}|x)p_{data}(x)}\left(\lambda(\sigma, \tilde{x} - x)\left\|s_\theta(\tilde{x}) + \frac{\tilde{x} - x}{\sigma^2}\right\|_2^2\right) \tag{16}$$

Using deep learning techniques, the neural network $s_\theta(\tilde{x})$ can be trained on an image dataset to establish the advanced score function (ADSF).

**Dataset**

We used CT images in NIH-AAPM-Mayo Clinic CT Grand Challenge dataset to train the advanced score function. The dataset consists of 2,378 CT images from 10 patients with a slice thickness of 3mm. We randomly divided the dataset into a training dataset and a test dataset. The training dataset contains 1,923 images from 8 patients, and the test dataset has 455 images from the remaining 2 patients.

**Neural Network**

The standard training process was performed for the training, validation, and testing. We established a ResNet neural network to learn the advanced score function (ADSF) from CT image dataset based on Eq. (16). ResNet has four residual blocks, followed by a convolution layer with 16 filters of 3×3 kernels, a convolution layer with 8 filters of 3×3 kernels, and finally a convolution layer with a single filter of 3×3 kernels to generate one feature map as the output. The first two residual blocks each have three sequential convolutional layers with 32 filters of 5×5 kernels and a residual connection. The last two residual blocks each have three sequential convolutional layers with 32 filters of 3×3 kernels and a residual connection. Every convolution layer is followed by a ReLU activation function, as shown in **Figure 4**.

**Data Training**

Noise levels of Gaussian mixture parameters in Eq. (16) were set to: 0.00, 0.01, 0.02, 0.03, 0.04, 0.05, 0.06, 0.07, 0.08, 0.09, 0.10, which corresponding probability distribution was the binomial distribution with 10 trials and the probability of success 0.5: 0.0010, 0.0098, 0.0439, 0.1172, 0.2051, 0.2461, 0.2051, 0.1172, 0.0439, 0.0098, 0.0010, respectively. The parameters of the convolution kernels were randomly initialized by a Gaussian distribution with a mean of zero and a standard deviation of 0.01. Using image patches of 32×32, ResNet is trained to optimize the kernels in the convolutional layers by minimizing the loss function according to formula Eq. (16) on a training dataset. The training procedure was programmed in Pytorch on a PC computer with an NVIDIA Titan XP GPU of 12 GB memory. The optimization was conducted using the ADAM algorithm with the exponential decay rate $\beta_1$=0.9 for the first moment estimates and $\beta_2$=0.999 for the second-moment estimates. The training proceeded at a learning rate of $10^{-4}$ over 1,000 epochs, and took approximately 24 hours. This training process showed excellent convergence and stability. The trained ResNet established the advanced score function for image reconstruction, which was used in the iterative formula Eq. (8) of the Methods section.

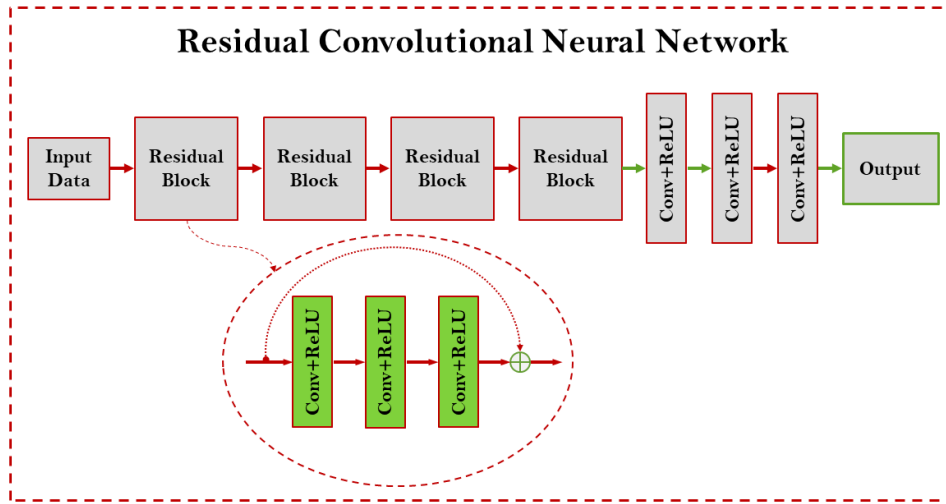

*Figure 4. Residual neural network architecture*

**Data availability**

CT dataset are publicly available online at https://www.aapm.org/grandchallenge/lowdosect/

**Code availability**

Source codes are publicly available.

**References**

[1]  T. M. Buzug, *Computed tomography : from photon statistics to modern cone-beam CT*, Berlin: Springer, 2008.

[2]  A. C. Kak, and M. Slaney, *Principles of computerized tomographic imaging*, Philadelphia: Society for Industrial and Applied Mathematics, 2001.

[3]  I. A. Elbakri, and J. A. Fessler, "Statistical image reconstruction for polyenergetic X-ray computed tomography," *IEEE Trans Med Imaging,* vol. 21, no. 2, pp. 89-99, Feb, 2002.

[4]  J. Tang, B. E. Nett, and G. H. Chen, "Performance comparison between total variation (TV)-based compressed sensing and statistical iterative reconstruction algorithms," *Phys Med Biol,* vol. 54, no. 19, pp. 5781-804, Oct 7, 2009.

[5]  K. Sauer, and C. Bouman, "A local update strategy for iterative reconstruction from projections," *IEEE Transactions on Signal Processing,* vol. 41, no. 2, pp. 534-548, 1993.